\documentclass[12pt]{article}
\catcode`\@=11
\@addtoreset{equation}{section}
\catcode`\@=12

\usepackage{amsbsy}
\newcommand{\B}{\mathbf B}
\newcommand{\A}{\mathbf A}
\newcommand{\abf}{\mathbf a}
\newcommand{\At}{\tilde \mathbf A}
\newcommand{\X}{\boldsymbol{\times}}
\newcommand{\x}{\bf x}
\newcommand{\D}{\mathbf D}
\newcommand{\W}{\mathbf W}
\newcommand{\om}{\boldsymbol{\omega}}
\newcommand{\del}{\boldsymbol{\nabla}}
\newcommand{\ru}{\hat \mathbf r}
\newcommand{\thetau}{\hat{\boldsymbol{\theta}}}
\newcommand{\phiu}{\hat{\boldsymbol{\phi}}}
\newcommand{\clqJM}{\mathbf C^{\lambda}_{qJM}}
\newcommand{\caqJM}{\phi^{\alpha}_{qJM}}
\newcommand{\tp}{({ \theta , \phi})}
\newcommand{\lJM}{\sum_{J,M, \lambda}}
\newcommand{\aJM}{\sum_{J,M ,\alpha}}
\newcommand{\alJM}{\sum_{J,M ,\lambda,\alpha}}
\newcommand{\rtp}{(r, \theta , \phi)}
\newcommand{\flqJM}{f^{\lambda}_{qJM}}
\newcommand{\flaqJM}{f^{\lambda \alpha}_{qJM}}
\newcommand{\blm}{b_{\lambda \mu}}
\newcommand{\bml}{b_{\mu \lambda}}
\newcommand{\bnm}{b_{\nu \mu}}
\newcommand{\fm}{f^{-}_{qJM}}
\newcommand{\fp}{f^{+}_{qJM}}
\newcommand{\fpm}{f^{\pm}_{qJM}}
\newcommand{\fmqJM}{f^{\mu}_{qJM}}
\newcommand{\fnqJM}{f^{\nu}_{qJM}}
\newcommand{\fo}{f^{0}_{qJM}}
\newcommand{\cpqJM}{\mathbf C^{+}_{qJM}}
\newcommand{\cmqJM}{\mathbf C^{-}_{qJM}}
\newcommand{\cpmqJM}{\mathbf C^{\pm}_{qJM}}
\newcommand{\cmuqJM}{\mathbf C^{\mu}_{qJM}}
\newcommand{\czqJM}{\mathbf C^{0}_{qJM}}
\newcommand{\jt}{\cal J }
\newcommand{\ap}{a_{+}}
\newcommand{\gpm}{g^{\pm}_{qJM}}
\newcommand{\gp}{g^{+}_{qJM}}
\newcommand{\am}{a_{-}}
\newcommand{\gm}{g^{-}_{qJM}}
\newcommand{\la}{\Lambda}

\newcommand{\bea}{\begin{eqnarray}}
\newcommand{\eea}{\end{eqnarray}}
\title{{\small\hfill IMSc/2003/04/04}\\ 
\textbf{Perturbation theory including topological degrees of freedom:
       Yang-Mills theory in three Euclidean dimensions }}
\author{E. Harikumar\footnote{hari@imsc.res.in},~
Indrajit Mitra\footnote{indrajit@imsc.res.in}~ and
H. S. Sharatchandra\footnote{sharat@imsc.res.in} \\\\
The Institute of Mathematical Sciences,\\ C.I.T. Campus, Taramani P.O.,\\
Chennai 600 113, India}
\date{}

\begin{document}
\maketitle

\begin{abstract}
  
  A method for systematically including topological degrees of freedom
  in perturbation theory is developed. This is not bound by the
  restrictions of semi-classical techniques. The Yang-Mills theory in
  three Euclidean dimensions is considered here. 
  A well-defined separation of the topological and the ``spin
  wave'' degrees of freedom is obtained, motivated by a singular gauge. This
  has ``photons'' distorting the spherically symmetric magnetic fields
  of Dirac monopoles, and massless charged vector bosons ``$\W$''
  scattering off the latter. It is explicitly shown that the Dirac
  string does not contribute. The mode of the charged vector bosons
  with total angular momentum $J=0$ provides precisely the core to
  give a finite energy to the monopole. The radial equation for $\W$
  is remarkably simplified and only two polarization states survive
  exactly for the anomalous magnetic moment required by the Yang-Mills
  interaction.

\end{abstract}
\newpage
\section{Introduction}
It is expected that the topological degrees of freedom such as
monopoles, instantons, $Z_N$-vortices etc.\ are responsible for crucial
properties such as confinement, deconfinement phase transition,
$\eta-\eta'$ mass difference etc. An example is provided by the
Georgi-Glashow model in $2+1$ dimensions. Classically the system may
be in the Higgs phase, but the topological degrees of freedom restore
confinement \cite{p}. A clean demonstration of this is possible
because there is a regime of parameters where semiclassical techniques
and the dilute gas approximation can be strictly justified.  This uses
the finite-energy, classically stable 't Hooft-Polyakov monopole
solution of the Euclidean theory. We have considered this example to
highlight the differences with the case of the $2+1$ dimensional
Yang-Mills theory which we address in this paper. In three Euclidean
dimensions, the theory does not have classically stable monopole
solutions.  Nevertheless, consider the 't Hooft-Polyakov ansatz
\cite{thp} for the non-Abelian gauge potential : 
\bea \label{thp}
A_i^a(x) = \epsilon _{iab} x^b \frac{1-K(r)}{r^2} 
\eea 
where $i=1,2,3$
labels the space index and $a,b=1,2,3$ label the group indices.  Here
$r=\sqrt{x^ax^a}$. As long as $K(r)=1+O(r^2)$ as $r\rightarrow 0$ and
$K(r)\rightarrow 0$ as $r\rightarrow \infty$, this configuration has a
finite (Euclidean) action. Moreover it has strong qualitative effects.
Consider a large Wilson loop
\bea
W [C] = P \exp (i \oint _C dx^i A_i^a(x) \sigma ^a/2)
\eea
(where $P$ stands for the path ordering along the loop $C$, and
$\sigma_a$ are the Pauli matrices). This configuration gives $ W [C] =
\exp(i \Omega /2) $ where $\Omega$ is the solid angle subtended by the
loop at the origin. Such configurations have the potential to disorder
and change the expectation value of $W[C]$ from the perimeter law to
the area law. The reason is that for a monopole near the plane of the
loop, $\Omega\sim 2\pi$ and $W[C]\sim -1$, so that a gas of
(anti-)monopoles can give coherent cancellations and area law.  A
suggestive calculation comes from the vortices in two (Euclidean)
dimensional Abelian Higgs model \cite{kogut}. In the dilute gas
approximation, with the (anti-)vortex having a chemical potential
$\mu$, we get
\bea
W[C] \sim \sum_{n=0}^{\infty} (-1)^n \frac{A^n}{n!} (e^{- \mu})^n 
= \exp (-e^{-\mu}A)\,.
\eea
(Here we have summed over vortices only, though anti-vortices are to
be included in an analogous manner.) $A$ is the area of the loop.The
entropy factor $A^n$ is there because each vortex can be located
anywhere inside the loop, and $(n!)^{-1}$ is the symmetry factor. As
each vortex contributes $-1$, we get the alternating sign factor
$(-1)^n$. Because of the coherent cancellation, $W[C]$ is much smaller
(follows area law) than what perturbative calculation would suggest.

Similar arguments are in operation in the three-dimensional
Georgi-Glashow model.  We may expect an analogous mechanism to be
responsible for confinement in the Yang-Mills theory in three
Euclidean dimensions too. 
Configurations  such as in (\ref{thp}) have finite action, and 
should be included in the sum over configurations,
\bea \label{fi}
Z = \int D \A ~  \exp \left (-\frac{1}{2e^2} \int d^3x \B^2(x) \right )
\eea
where $\B^a(x)= \nabla \X \A^a (x) - \frac{1}{2} \epsilon^{abc} \A^b
(x) \X \A^c (x)$ is the non-Abelian magnetic field.  But the
conventional semiclassical techniques are not applicable as there is 
no stable classical monopole solution. Indeed, the
situation is more demanding.
The form factor $K$ in Eq.\ (\ref{thp}) is actually a function
of $r/a$ where $a$ is the length scale associated with the monopole. 
Then the action must be proportional to $a^{-1}$ due to dimensional reasons.
(This is the reason for the classical instability of
this configuration: the action is reduced as the monopole expands.) As
a consequence, monopoles of very large size are very light, and
therefore can proliferate. The situation is exactly the opposite of
one where the dilute gas approximation can be justified.  We need a way
to handle large, light and overlapping monopole configurations. In
Ref. \cite{hss} it was proposed to do this by summing over the
``topological centres'' of monopoles with a constraint related to the
size of the monopole. The semiclassical technique can be modified in
principle to carry out this computation. But explicit computations are
technically difficult because the dilute gas approximation is not
valid, and the energetics of the multi-monopole configurations has to
be addressed first.

In this and succeeding papers, we develop a new approach for
systematically including topological degrees of freedom within the
ambit of renormalized perturbation theory. Our technique does not
rely on the existence of stable classical solutions.

This paper is organized as follows: In Sec.\ 2, we obtain a well 
defined separation of the ``topological'' and the ``spin wave'' 
degrees of freedom. We argue that the topological degrees 
survive integration over the spin wave degrees, even though we 
are not expanding about an extremum of the action.
In Sec.\ 3, we analyse the scattering of a charged vector boson
off a Dirac monopole. We show that there is a remarkable 
simplification of the radial equation  and one polarization 
state decouples only when the anomalous
magnetic moment is two, exactly as required by the Yang-Mills
interaction. In Sec.\ 4, we obtain the zero modes of the 
Hamiltonian and explain why one polarization
state decouples. In Sec.\ 5, we re-express the functional integral
using the new variables to bring it to a form where 
perturbative calculation including the topological degree can be
carried out (in the case of one monopole). In Sec.\ 6, we present our
conclusions.
\section{Separation of ``spin waves'' and ``topological'' degrees of freedom in the Yang-Mills
theory} \label{sep}
In Ref.\ \cite{topo}, we have obtained a gauge-invariant
characterization of all possible topological configurations of the
Yang-Mills theory. For this, we considered the eigenvalue equation
\bea
S^{ab}(x) \ \xi_b^A(x)=\lambda^A(x) \ \xi_a^A(x),~~ A=1,2,3
\eea
where $S_{ab}$ is the $3\times 3$ symmetric matrix
\bea
S^{ab}(x)=\B^a(x) \cdot \B^b(x)\,,
\eea
transforming covariantly (as the adjoint representation) under the
$SO(3)$ local gauge group.  The three normalized eigenvectors
$\xi_a^A(x)$ form an orthogonal matrix (field). Points and lines of
singularities of these eigenvectors (or equivalently, loci of
degeneracies of the eigenvalues $\lambda^A(x)$) characterize the
topological configurations. In particular, it locates the ``topological 
centre'' of a monopole configuration such as in Eq. (\ref{thp}) in a
gauge-invariant manner. This is of great advantage to characterize a
dense gas of monopoles (and other topological objects) especially when
semiclassical techniques are inapplicable.  We may use the orthogonal
matrix $\xi_a^A(x)$ to formally perform a gauge transformation on the
monopole configuration (\ref{thp}). The transformed gauge potential
$\abf^A(x)$ is
\bea
\abf^A ={\tilde \A}^A+{\om}^A
\eea
where
\bea
{\tilde \A}^A=\xi_{a}^A \A^a
\eea
and
\bea
\om^A=\frac{1}{2}\epsilon^{ABC}\xi_{a}^B\nabla\xi_{a}^C.
\eea
We obtain
\bea
\label{2.6}
\abf^1= \phiu \frac{K(r)}{r}, ~~
\abf^2= - \thetau \frac{K(r)}{r}, ~~
\abf^3= - \phiu \frac{\cot \theta }{r}
\eea
where ($\ru$, $\thetau$, $\phiu$) are the unit vectors of the
spherical coordinates ($r$, $\theta$, $\phi$).  $\abf^3$ is precisely
the point Dirac monopole, with two Dirac strings along $\pm z$
directions.  (We have relabelled the indices $A=1,2,3$, as compared to
Ref.\ \cite{topo}, to have the Dirac monopole for $A=3$.) Even though
the configuration  Eq. (\ref{thp}) is non-singular, we have now got a
singular configuration because $\xi_a^A(x)$ is itself singular on the
$z$-axis (the angle $\phi$ being undefined at $\theta=0$ and
$\theta=\pi$). So $\xi_a^A(x)$ is not strictly an allowed local gauge
transformation. Indeed the non-Abelian magnetic field for the gauge
potential $\om^A$ is not zero; rather, $\B[\om]=\mbox{Dirac string
  contribution}$. Nevertheless, this singular gauge transformation
highlights the topological objects and can be handled without
problems, as will be shown below.

First we show that the Dirac string singularity does not contribute to the action.
In spite of the singularity on the $z$-axis, we have 
$\xi _a^A (x) \xi _b^A (x) =\delta_{ab}$ everywhere. Therefore  
\bea
\B ^a (x) \cdot \B ^a (x) = (\xi _a^A (x) \B ^a (x)) \cdot
(\xi _b^A (x) \B ^b (x))\,.
\eea
Now,
\bea
\xi _a^A  \B ^a  &=& \del \X ( \xi _a^A \A ^a) -\del \xi _a^A \X \A ^a
- \frac{1}{2} \epsilon^{abc} \xi _a^A \A ^b \X \A ^c \, \nonumber \\
 &=& \del \X \At ^A - \epsilon^{ABC} \om ^B \X \At ^C
  -\frac{1}{2} \epsilon^{ABC} \At^B \X \At^C \, \nonumber \\
      &=&\B^A [\om + \At] - \B^A [\om] =\B^A [\abf ] - \B^A [\om]\,.
\eea
Here $\B^A [\abf ]$ stands for the non-Abelian magnetic field for the
gauge potential $\abf$.  Note that these operations are valid even if
$\xi _a^A$ is singular.  Subtracting by $\B[\om]$ precisely removes
the singular contribution from the Abelian curl of $\abf$, which is
the Dirac string magnetic field. This was to be expected because $\xi
_a^A (x) \B ^a $  in Eq.\ (\ref{fi}) is finite everywhere. Therefore we may simply replace
$\B^a \cdot \B^a $ by $\B^A [\abf ] \cdot \B^A [\abf]$ with the
understanding that the Dirac string should be ignored in computing
$\del \X \abf^3$.  With this convention,
\bea
\label{2.9}
\B^3 [\abf ]=\del \X \abf^3 -\abf^1 \X \abf^2\,.
\eea
$\del \X \abf^3 $ is the magnetic field of the Dirac monopole (sans
the string) and is singular as $\ru /r^2 $ at the monopole centre.
This singularity is precisely cancelled by the non-Abelian interaction
term $-\abf^1 \X \abf^2 = - \ru (K(r))^2/r^2 $ since $K(r)= 1+O(r^2)$
for $r\rightarrow 0$.

To use the Dirac potential for the monopole,
a slightly different singular gauge transformation will be used in the present work,
namely, $\xi_a^1=\cos\phi\,\hat\theta_a-\sin\phi\,\hat\phi^a$,
$\xi_a^2=\sin\phi\,\hat\theta_a+\cos\phi\,\hat\phi^a$, and $\xi_a^3=\hat x_a$.
The potentials obtained on performing this transformation on the configuration
(\ref{thp}) may be expressed as
\bea
\label{ourpot}
\abf^1+i\abf^2=(\phiu - i \thetau)\frac{K(r)}{r}e^{i\phi}, ~~~
\abf^3=\phiu\frac{1-\cos\theta}{r\sin\theta}\,.
\eea
The cancellation described after Eq.\ (\ref{2.9}), of course, still takes place.
It may be noted that the potential given in Eq.\ (\ref{ourpot}) is related to
the potential given in Eq.\ (\ref{2.6}) by a $U(1)$ gauge transformation.

We advocate this picture obtained in the singular gauge to obtain a
well-defined separation of the topological degrees of freedom and the
`spin waves', and as a means of including the topological degrees
systematically in renormalized perturbation theory. This method can
handle all topological degrees such as vortices, half-monopoles,
half-vortices \cite{half} etc. But in this paper we consider only
monopole configurations such as in Eq.\ (\ref{thp}).

The general ansatz for $\abf^3 $  is
$\abf^3= \A + \abf$ with
\bea
\label{vecpot}
\A(x)=\sum_{n} q_n \phiu_n \frac{1-\cos \theta_n}{|\x-\x_n|\sin\theta_n}\,.
\eea
$\A(x)$ is the superposition of Dirac potentials of charge $q_n=\pm 1$ 
(in units of $1/e$) located at $\x=\x_n$, 
$n=1,2,3,\cdots$.
Also, $(r_n$, $\theta_n$, $\phi_n)$ are the spherical coordinates centred at $\x=\x_n$, and 
$(\ru_n$, $\thetau_n$, $\phiu_n)$ are the
corresponding unit vectors. The linear superposition of the (spherically symmetric) magnetic 
fields of the monopoles is distorted by the magnetic fields from the potential $\abf$ representing
the `spin wave' degrees of freedom. We will refer to $\abf$ as the {\it photon}. We consider the 
linear combination $\W = (\abf^1+i\abf^2)/ \sqrt 2 $ ~ (and 
$\W^{\star} = (\abf^1-i\abf^2)/ \sqrt 2 $), and refer
to it as the {\it charged vector boson $ $}. The action $S= \int d^3x\, \B^2/2$
then becomes
\bea \label{ac}
S = \int d^{3}x  \left ( \frac{1}{2} ( \sum q_n \frac{\x - \x_n} {| \x - \x_n|^3}
+ \nabla \X \abf + i \W^{\star}  \X \W )^2
+ | \D [\A + \abf ] \X  \W |^2 \right )  
\eea
where we have ignored the Dirac string contribution as was shown before. Here
\bea
\D [\A]= \del -i\A
\eea
is the Abelian covariant derivative. 

Note that the non-Abelian interaction term from $i \W^{\star} \X \W $ has
the physical meaning of an anomalous magnetic moment $g=2$ for the
charged vector boson.  This interaction turns out to provide
remarkable simplification in the scattering of the charged vector
boson off the magnetic monopole, as will be discussed in Sec.\ 
\ref{scat}.  Also the $J=0$ mode of this interaction provides the core
to regulate the energy of the Dirac monopole in the ultraviolet.

The action is finite only when the singular energy of the point Dirac
monopole is cancelled by the cloud of charged vector bosons around it:
$i \W^{\star} \X \W $ should behave like
\bea
-q_n \frac{\x - \x_n} {| \x - \x_n|^3}+\mbox{finite terms} ~~~ {\rm for} ~ \x \rightarrow \x_n . 
\eea
This can be obtained by
\bea
\W(x) \rightarrow (\phiu_n - i \thetau_n)\frac{e^{i\phi_n}}{\sqrt 2 |\x-\x_n|} + O(| \x - \x_n|).
\eea
Note that this is in accord with the behaviour of the potential given in Eq.\ (\ref{ourpot})
near the origin.
As will be seen in Sec.\ \ref{funi},
the $J=0$ mode of the
$\W$ boson also has precisely this behaviour.

We have therefore arrived at the picture of {\it a cloud of massless
  charged vector bosons regularizing the energy of Dirac monopoles,
  and photons distorting the (spherically symmetric) magnetic fields
  of the monopoles.} Even though we do not have the stable
  finite-energy classical solutions, we can handle this action
  systematically by renormalized perturbation theory. In this way we can
  include the topological degrees in the usual perturbation theory.
  
  In Eq.\ (\ref{ac}), it might appear that we have expanded the action
  about a background (i.e. Dirac monopole) which is not a solution of the classical
  equations of motion. If such were the case, an integration over
  the quantum fluctuations would wash out the effects of the background,
  as will be
  discussed now. Consider for illustration a (Euclidean) scalar field
  theory
\bea
Z = \int D \phi ~ e^{-S[ \phi]}
\eea
where the action is expanded about a background $\phi_0$. If $\phi_0$
is not a solution of the classical equation of motion,
terms linear in the fluctuation
$\chi=\phi-\phi_0$ will be present: $S[ \phi]=S[ \phi _0]+s[\chi ,
\phi _0]$ where $S[ \phi _0]$ is the ``classical action'' and
\bea
s[\chi , \phi _0]  =  \chi \cdot \frac{ \delta S[ \phi _0]}{ \delta \phi _0 }
+\frac{1}{2} \chi \cdot \frac{ \delta ^2 S[ \phi _0]}
{ \delta \phi _0 \delta \phi _0} \cdot \chi + 
({\rm higher~ orders~ in~ field~} \chi)\,.
\eea
In a free theory the higher order terms are absent. In this case
integration over the fluctuations ``cancels'' the ``classical ''
contribution $S[ \phi _0]$. In an interacting theory, we have to
first remove the linear term in $s[\chi , \phi _0]$ by a shift of the field $\chi$.
This shift again cancels the ``classical '' term.

We now argue that the situation is different in our case. Consider 
the theory \cite{manu}
\bea
Z = \int DH_{\mu \nu} DA_{\mu} ~ \exp \left (-\frac{1}{4} 
(\partial_{\mu} A_{\nu}-\partial_\nu A_{\mu}
+H_{\mu \nu})^2-s[H_{\mu \nu}] \right )\,.
\eea
$H_{\mu\nu}$ describes (a part of) the field carrying magnetic charge. We first integrate
over $A_\mu$ (say choosing the Landau gauge):
\bea
Z = \int DH_{\mu \nu}\exp(-\frac{1}{4}H_{\mu \nu}^2 +\frac{1}{2} \partial_\nu H_{\mu \nu} \cdot 
\partial^{-2} \cdot \partial_\rho H_{\mu \rho}-s[H_{\mu \nu}])\,.
\eea
Now $H_{\mu\nu}$ terms are not cancelled out completely:
\bea
-\frac{1}{4}H_{\mu \rho}(\delta_{\rho \sigma}-2 \partial_{\rho} \partial_\sigma/\partial^{2})
H_{\mu \sigma}\,.
\eea
The reason is clear: the part of $H_{\mu\nu}$ which is of the form
$\partial_\mu a_\nu -\partial_\nu a_\mu$ is removed by the shift
$A_\mu\rightarrow A_\mu+a_\mu$. The degrees of freedom in $H_{\mu\nu}$
that are not represented by $a_\mu$ give a non-zero contribution.

This analysis is directly relevant to us. In Eq.\ (\ref{ac}), the magnetic field of a
Dirac monopole is a pure gradient, while that of a `photon' is a pure
curl. Therefore the contribution of the monopole survives the
integration over the gauge potential $\abf$. Indeed, the cross term of
the monopole field and $\abf $ is a total derivative and drops out of
the action, as in the case of an expansion about an extrema of the
action. Therefore the topological degrees survive integration over 
the spin wave fluctuations.

\section{A massless charged vector boson scattering off a Dirac monopole}
\label{scat}
In this section we obtain the eigenfunctions for a charged vector 
boson in the background of a Dirac monopole.  The results will be used 
in Sec.\ \ref{funi} to express 
the functional integral in the new modes.  The interaction of a spin-one 
particle with a magnetic monopole has been analyzed before in Ref.\ \cite{wu}. 
We consider here a massless vector boson in contrast to the massive case which 
they have addressed. Our analysis and results turn out to be considerably different.
We find that for precisely the anomalous magnetic moment $g=2$ as required by the
Yang-Mills interaction Eq.(\ref{ac}) there is a dramatic simplification of the radial equations.
Also, only in this case, one of the three polarization states of the charged vector
boson decouples in correspondence with the free theory.

The eigenvalue equation we consider is
$H \W = k^2 \W$, where
\bea \label{h}
H= (\D \X \D - iq \frac{\ru}{r^2}) \X  
\eea
where $k^2$ is the energy eigenvalue. Now $\D$ stands for the covariant derivative with 
the Dirac potential of a monopole at the origin. We may expand $\W$ as follows:
\bea \label{vh}
\W \rtp = \lJM \flqJM (r) \clqJM \tp
\eea
where $\flqJM (r)$ is a function of $r$ only, and the dependence on $\theta$ and $\phi$ 
is contained in
$\clqJM$, which are the monopole vector spherical harmonics for the
monopole of strength $q$. It may be noted that {\it $\clqJM$ is defined with
a factor of $1/r$}: see Eq.\ (\ref{ortho}). Here $J(J+1)$ and $M$ refer to the
eigenvalues of $\bf J^2$ and $J_z$ where $\bf J$ is the total angular
momentum $\bf L+\bf S$.  The orbital angular momentum $\bf L$ of a
charged particle in the magnetic field of a Dirac monopole is given by
$ {\bf L}= -i {\bf r} \X \D -q \ru $. The extra term $-q \ru$ is the
well known angular momentum carried by the crossed electric and
magnetic fields, and plays a significant role in the interaction of
the charged particle with the monopole. $\bf S$ is the spin operator
for the vector boson: $(S^k)_{ij}=-i\epsilon_{ijk}$, ($i,j,k=1,2,3$),
acting on the three components of the vector $\W $.  We will adopt the
approach of E. J. Weinberg \cite{wein}, where the multiplets for
given $J$ and $M$ are labelled by $\lambda=0,\pm$ according as the
eigenvalues of ${\ru} \cdot \bf{S} $ are $0, \pm 1$. We have
\bea
\D \X \clqJM = \frac{i}{r} \sum_{\mu} \blm \cmuqJM
\eea
where the only non-vanishing $\blm$ are
\bea
b_{0,\pm}= -  b_{\pm , 0}= \pm  a_{\pm}
\eea
where $a_{\pm}=\sqrt{({\jt}^2 \pm q)/2}$ and  ${\jt} =\sqrt{J(J+1)-q^2}$. 

Note that $\blm$ is antisymmetric in its indices. This corrects an error in Ref.\ \cite{wein}.
(In Eq.\ (3.10) of this reference, 
$r \int d \Omega ~ 
{\mathbf C^{0\star}_{qJ^{'}M^{'}}}
\cdot {\tilde \D} \X \cpmqJM$ is equal to 
 $r \int d \Omega ~ ({\tilde \D} \X {\mathbf C^{0}_{qJ^{'}M^{'}}})^\star
\cdot  \cpmqJM $, and not its negative. The reason is that
not only the integration by parts gives a negative sign, but also the interchange of 
$\mathbf C^{0 \star}_{qJM}$ and ${\tilde \D}$
in the cross-product.)

The following discussion will make it clear that the action of $H$ does not change 
the $J$ and $M$ values for 
the terms in Eq.\ (\ref{vh}). Therefore, henceforth in this section, we take $\W$ to be 
an eigenstate of $J^2$ and $J_z$, i.e., we consider  
$\W = \sum_\lambda\flqJM  \clqJM $.

Using 
\bea
\label{rcc}
\ru \X \clqJM = -i \lambda \clqJM ,     
\eea
we then get
\bea
\D \X \W = \sum_\lambda\left(-i \lambda \frac{d\flqJM}{dr}+ \frac{i}{r}\sum_\mu \fmqJM \bml \right )\clqJM
\eea
and 
\bea
\D \X ( \D \X \W) & =& \sum_\lambda \Bigg(-\lambda^2 \frac{ d^2 \flqJM}{dr^2}
+\sum_\mu \frac{(\lambda+\mu)}{r}\frac{d\fmqJM}{dr}\bml
- \frac{\lambda}{r^2}\sum_\mu \fmqJM \bml \nonumber\\
&&-\frac{1}{r^2}\sum_{\mu,\nu} \fnqJM \bnm\bml\Bigg)\clqJM~
\eea
We put this expression in the eigenvalue equation, and also make use of Eq.\ (\ref{rcc}).
This leads to, for $\lambda=0,+,-$ respectively:
\bea
-\frac{d}{dr} (\ap \fp + \am \fm) + {\jt} ^2 \frac{\fo}{r} &=&
k^2 r \fo \,                                                   \label{111a} \\
-\frac{d^2}{dr^2} \fp + \ap  \frac{d}{dr} \frac{\fo}{r}
+ \frac{\ap}{r^2} (\ap \fp - \am \fm) + [-q \frac{ \fp}{r^2}] &=&
k^2  \fp \,                                                     \label{111b}\\
-\frac{d^2}{dr^2} \fm + \am  \frac{d}{dr} \frac{\fo}{r}
- \frac{\am}{r^2} (\ap \fp - \am \fm) - [-q \frac{ \fm}{r^2}] &=&
k^2  \fm                                                        \label{111c}
\eea
We have bracketted the contribution from the anomalous magnetic moment to highlight an
important cancellation. Using the combinations
\bea
\gpm = \ap \fp \pm \am \fm
\eea
we get
\bea
-\frac{d}{dr} \gp  + {\jt}^2 \frac{\fo}{r} &=&
k^2 r \fo                                                    \label{222a} \\
-\frac{d^2}{dr^2} \gp + {\jt}^2  \frac{d}{dr} \frac{\fo}{r}
+ q \frac{ \gm}{r^2} + [-q \frac{ \gm}{r^2}] &=&
k^2  \gp                                                      \label{222b} \\
-\frac{d^2}{dr^2} \gm + q  \frac{d}{dr} \frac{\fo}{r}
+ {\jt}^2 \frac{ \gm}{r^2} - [q \frac{ \gp}{r^2}] &=&
k^2  \gm                                                     \label{222c} 
\eea
Precisely due to the anomalous magnetic moment term, $\gm$ drops out
in Eq.\ (\ref{222b}). This simplifies the equations
dramatically, as shown below. Comparing the derivative (with respect
to $r$) of Eq.\ (\ref{222a}) with Eq.\ (\ref{222b}), we obtain
\bea
\gp =  \frac{d}{dr}( r \fo)\label{gplus}
\eea
Plugging this back into Eq.\ (\ref{222a}), we get a simple second-order equation for $\fo$:
\bea
\frac{d^2}{dr^2} (r \fo) - {\jt}^2   \frac{\fo}{r}
+k^2  r \fo = 0 
\eea
Thus $\fo$ satisfies precisely the radial equation for a free particle
with the eigenvalue $l(l+1)$ of $\bf L^2$ replaced by ${\jt}^2$. If
the anomalous magnetic moment term were not included (or if it had any
value other than $g=2$), the equation satisfied by $\fo$ would be of
third order and contrived.

We may write the equation satisfied by $\gm$ as
\bea
\label{333}
\frac{d^2}{dr^2} \gm - {\jt}^2 \frac{\gm}{r^2} +k^2\gm = -2q \frac{\fo}{r^2}
\eea

Let us now discuss the relevant solutions to these equations.
\begin{enumerate}
     \item{Choose $\fo=0$; therefore from Eq.\ (\ref{gplus}), $\gp=0$. We
          also see from Eq. (\ref{333}) that $\gm/r$ then satisfies the equation for
          the radial part of the free particle wavefunction  (with
          $l(l+1)\rightarrow {\jt}^2$). The acceptable solution is
\bea
\label{444}
\frac{\gm}{r}=\sqrt{\frac{\pi}{2kr}}J_{\sqrt{\frac{1}{4}+{\jt}^2}}(kr)
\eea
      
         (where $J$ denotes the Bessel function) 
	 with the behaviour $\gm\sim r^{\frac{1}{2}+\sqrt{\frac{1}{4}+{\jt}^2}}$ 
	 near the origin. Note that the order of the Bessel function is not half-integral,
         in contrast to the case of the usual free-particle radial equation.
         
         In this case, we considered the general solution of the
         homogeneous equation for $\gm$.
	 Next we consider the particular solution for the inhomogeneous equation.}
     \item {The acceptable solution for $r\fo$ is the same as for
         $\gm$ in Eq.\ (\ref{444}). The corresponding $\gpm$ are 
         obtained by solving the inhomogeneous equations (\ref{gplus})
	 and (\ref{333}).}
\end{enumerate}

We notice that there are only two linearly independent set of
solutions, although there are three `polarization states' $\lambda=0,\pm$
for the vector boson. The reason is analogous to the case of free
massless vector bosons.  This will be demonstrated in Sec.\ \ref{ker}.

In solving the eigenvalue equation, we treated all values of $J$ and
$M$ on equal footing.  However, the following cases are special.

\begin{enumerate}
\item{$J=q-1$: There is only one multiplet $\lambda=+$. 
The mixing matrix $ $ is now formally zero.
So putting $a_\pm=0$ in Eq.\ (\ref{111b}), we get the equation
\bea
\label{555}
\frac{d^2}{dr^2} \fp +q \frac{ \fp}{r^2} +k^2  \fp = 0
\eea
for the only non-vanishing radial wavefunction $\fp$. As $q>0$ (the equations of Ref.\ \cite{wein}
are valid for $q\ge 0$ only), this effectively has a centrifugal attraction in the place of
centrifugal repulsion.  In Sec.\ \ref{funi}, we use a different basis for this mode.}
\item{$J=q$: There are only two multiplets, corresponding to $\lambda=0, +$. In this case 
$a_+=\sqrt q$ and $a_-=0$. We can recover the relevant equation for this case by
simply putting these values in the equations (\ref{111a}) and (\ref{111b}):
\bea
\frac{d}{dr} \fp - \sqrt q \frac{\fo}{r} +\frac{k^2 r}{\sqrt q}\fo = 0 \,  \nonumber \\
\frac{d^2}{dr^2} \fp - \sqrt q \frac{d}{dr} \frac{\fo}{r} +k^2
\fp = 0 
\eea
We have only one acceptable solution for $\fo$ and $\fp$.}
\end{enumerate}
\section{Kernel of the Hamiltonian} \label{ker}
We now obtain the modes of the Hamiltonian with zero eigenvalue. These
are relevant for the gauge-fixing condition \cite{p} as will be
considered later.

We first show that $\W=\D \la$ for any complex function $\la(x)$ is a
zero mode . We have
\bea
 \left (\D \X \D - iq \frac{\ru}{r^2} \right ) \X \D \la 
&=& \D \X \left (- iq \frac{\ru}{r^2} \la \right )
- iq \frac{\ru}{r^2} \X \D \la \,  \nonumber \\
&=& \del \X  \left (- iq \frac{\ru}{r^2}\right ) \la
+ iq \frac{\ru}{r^2} \X \D \la - iq \frac{\ru}{r^2} \X \D \la
\eea
which is zero because the magnetic field of a monopole is the gradient
of a scalar potential.

It is easy to check that this exhausts all modes of zero energy by
using an expansion in the monopole vector spherical harmonics as in
Sec.\ \ref{scat}. We just need to consider Eqs.\
(\ref{222a})-(\ref{222c}) with $k=0$. The equations may be written as
\bea
\frac{d \gp}{dr}-{\jt}^2 \frac{\fo}{r} &=&0 \\
\frac{d^2}{dr^2} ( \frac{q}{{\jt}^2} \gp - \gm ) -  \frac{{\jt}^2}{r^2}
( \frac{q}{{\jt}^2} \gp - \gm ) &=& 0         \label{4.2} 
\eea
The non-zero solutions of Eq.\ (\ref{4.2}) are $r^{\alpha_\pm}$ (for all $r$) with
$\alpha_\pm =\frac{1}{2}\pm\sqrt{\frac{1}{4}+{\jt}^2}$. They do not
give normalizable solution. Therefore we need to consider only the
trivial solution of Eq.\ (\ref{4.2}), namely, $\gm=({q}/{{\jt}^2}) \gp $. In this case,
substituting $f^0$ and $\gm$ as given by these equations in favour of
$\gp$ into Eq.\ (\ref{vh}), we get
\bea
\W &=& \lJM \left ( \fo \czqJM + \frac{1}{2a_{+}} (\gp + \gm) \cpqJM
+ \frac{1}{2a_{-}} (\gp - \gm) \cmqJM \right ) \,  \nonumber \\
&=& \frac{1}{{\jt}^2}\lJM \left [\frac{d \gp}{dr} \ru Y_{qJM} +
\gp \left ( \frac{ {\jt}^2 +q}{2 \ap} \cpqJM
+  \frac{ {\jt}^2 - q}{2 \am} \cmqJM \right) \right ] \, \nonumber \\
&=& \frac{1}{{\jt}^2} \lJM \left ( \ru \frac{d \gp}{dr} + \gp \D \right ) Y_{qJM}
= \D \la
\eea
where $\la= \lJM \gp Y_{qJM}/{\jt}^2$. 
Here we used the expressions for $\clqJM$ given in Eqs.\ (3.6) of Ref.\ \cite{wein}.
This confirms that the only 
zero-energy solutions are of the form $ \D \la$.

Now the eigenfunctions for $k \neq 0$ are orthogonal to the
eigenfunctions for $k=0$.  Therefore $\int d^3x\,(\D\la)^\star\cdot\W=0 $ for arbitrary
$\la (x)$, so that (on integration by parts) 
\bea
\D\cdot\W=0.
\eea
It is this
constraint which leaves only two independent polarization states for
the eigenfunctions.

It is interesting that only by including the anomalous magnetic moment
interaction (as required by the Yang-Mills interaction) we get these zero
modes. We now show that there are no zero modes without this term.
In this case the zero modes are the solutions of
\bea
\D \X ( \D \X \W)  = 0
\eea
We may first solve $\D \X \bf V = 0$ (where $\bf V =\D \X \W$).
Expanding $\bf V $ in vector harmonics, as in Eq.\ (\ref{vh}), we get the
following equation for the radial part:
\bea
-\lambda \frac{ d \flqJM}{dr}+ {\sum_{\mu}}\frac{\fmqJM}{r} \bml = 0.
\eea
This gives
\bea
\gm &=& 0 \,  \nonumber \\
\frac{d \gp}{dr} &=& {\jt}^2  \frac{ \fo}{r} \,  \nonumber \\
\frac{d \gm}{dr} &=& q  \frac{ \fo}{r} 
\eea
Therefore $\gm=\fo=0$ and $\gp(r)$ can at best be a constant. Thus the
most general solution is $\fpm(r)=c/a_\pm$, where $c$ is arbitrary.
Thus for the zero modes, $\bf V$ is $1/r$ times a function of $\theta$ and $\phi$.
This is not
normalizable, and hence there are no acceptable zero modes.

It has been noted many times in literature \cite{g2} that the natural
value for the gyromagnetic ratio of an elementary charged particle
coupling to the electromagnetic field is $g=2$. Many reasons suggest
this. It has been observed in \cite{jack} that for the charged vector
boson interacting with the photon, only for $g=2$ there is a local
gauge invariance $\W \rightarrow \W+ \D \la$. This is a generalization
of the invariance in the free theory which implies that one polarization
state is unphysical. (This is part of the
non- Abelian gauge invariance for us.)

\section{Functional integral in terms of the new modes} \label{funi}
In this section we express the functional integral given by Eqs.\ 
(\ref{fi}) and (\ref{ac}) in terms of the new modes.
Henceforth $q=1$ only.
We will separate the $J=0$ mode  as it plays a
special role: $\W=\W_0+\W_1$, where,
\bea \label{wo}
\W_{0} \rtp = w(r) {\mathbf C^{+}_{100}} \tp
\eea
is the $J=0$ mode and
\bea \label{w1}
\W_{1} \rtp =\alJM \int_{-\infty}^{\infty} dk ~ \caqJM (k) \flaqJM (kr) \clqJM \tp
\eea
involves all higher $J$ modes. Here $\caqJM (k), \alpha=1,2$
represents the amplitudes for the two linearly independent solutions
$f^{\lambda \alpha}_{qJM} (kr),~ \alpha=1,2$ of the radial
equations when $J > 1$. For $J = 1$ there is only one solution, and
$\alpha=1$ only in this case.  The orthonormality conditions are
\bea
\label{ortho}
\int d \Omega ~ {\mathbf C^{\lambda \star}_{qJM}} \tp \cdot 
{\mathbf C^{\mu}_{qJ^{'}M^{'}}}\tp= \frac{1}{r^2}
\delta_{JJ^{'}} \delta_{MM^{'}} \delta_{\lambda \mu}
\eea
and
\bea
\label{5.4}
\sum_{\lambda} \int_0^{\infty} d r {f^{\lambda \alpha \star}_{qJM}}(kr)
{f^{\lambda \beta}_{qJM}}(k^{'}r)
=\delta (k-k') \delta _{ \alpha  \beta}.
\eea
Note that the vector monopole harmonics $\clqJM$ are defined with a
$r^{-1}$ factor \cite{wein}. As a result, the measure in Eq.\ (\ref{5.4})
is $dr$ and not $r^2 dr$.

The vector harmonic for $J = 0$ is special. It is both (covariant)
divergence and curl free. The expression for this harmonic is
\bea 
{\mathbf C^{+}_{100}} \tp = \frac{\phiu -i\thetau}{\sqrt{8 \pi} r} e^{i\phi}
\eea
[This can be explicitly checked. In terms of the
components of a transverse vector $\W= W_{\theta} \thetau +W_{\phi} \phiu$,
the equation $\D \cdot \W =0$ reads
\bea
\frac{\partial}{\partial\theta}(\sin\theta W_\theta)
+\frac{\partial}{\partial\phi}W_\phi
-i(1-\cos\theta)W_\phi=0,
\eea
while
$\D \X \W =0$ reads
\bea
\frac{\partial}{\partial\theta}(\sin\theta W_\phi)
-\frac{\partial}{\partial\phi}W_\theta
+i(1-\cos\theta)W_\theta=0,\nonumber\\
\frac{\partial}{\partial r}(rW_\phi)=0,~~~
\frac{\partial}{\partial r}(rW_\theta)=0.
\eea
Note that our definition of the components $W_\theta$ and $W_\phi$
differ from those of Ref.\ \cite{wein}.]
Therefore $i\W_0^{\star} \X \W_0= -\ru |w(r)|^2/4\pi r^2$.
This is precisely of the type to give a form factor to the point
Dirac monopole. We get
\bea \label{formfactor}
\frac{\ru}{r^2} + i \W^{\star} \X \W =
\frac{\ru}{r^2} \left ( 1 - \frac{|w(r)|^2}{4 \pi} \right )
+i (\W_{0}^{\star}  \X \W_{1} + \W_{1}^{\star}  \X {\W}_{0}
+ \W_{1}^{\star}  \X {\W}_{1})
\eea
If in addition $w(r)$ satisfies the boundary condition
\bea
\frac{|w(r)|^2}{4\pi} = 1+O(r^2), ~~~r \rightarrow 0,
\eea
the action is finite. Thus the $J=0$ mode of the $\W$ boson can 
regularize the energy of the Dirac monopole as a consequence of its anomalous 
magnetic moment. The boundary condition $ |w(r)|=\sqrt{4\pi}$
is necessary to relate (through a singular
gauge transformation) our singular fields to non-singular Yang-Mills 
potentials as in the t'Hooft -Polyakov ansatz  Eq.\ (\ref{thp}), as noted
in Sec.\ \ref{sep}.

For $J=0$, the (allowed) eigenfunctions of the radial equation (\ref{555})
has the behaviour $\fp \sim r^{(1\pm i\sqrt 3)/2}$
near the origin for any $k$. As we want the boundary 
condition $ |w(r)|=\sqrt{4\pi}$, we will not use this complete set 
of radial functions for the 
$J=0$ mode. Instead, we consider the eigenfunctions of the Hamiltonian
$\D \X \D \X $ ~without the anomalous magnetic moment term. For the ansatz 
(\ref{wo}), the eigenvalue equation is now simply
\bea 
-\frac{d^2 w}{dr^2} =k^2 w
\eea
since $\D\X \mathbf C^{+}_{100}$ is zero. 
Now the radial wave function $w(r)$ is a linear combination of the Fourier
modes $\exp(ikr)$ and hence not required to vanish at $r=0$. 

Using the orthonormality
of $\clqJM$, we have
\bea 
\int d \Omega ~ \W_{0}^{\star} \cdot \ru \X {\W}_{1} =0
\eea
because $\ru\X$ does not change the $J,M$ values when
acting on $\clqJM$. Therefore Eq.\ (\ref{formfactor}) gives
\bea \label{5.10}
\frac{1}{2} \int d^{3} x  ( \frac{\ru}{r^2} + i \W^{\star}  \X \W )^2
&=&\int dr  \frac{2 \pi}{r^2} \left ( 1 - \frac{|w(r)|^2}{4 \pi} \right )^2 \nonumber\\
&&-i \int d^{3} x \left ( 1 - \frac{|w(r)|^2}{4 \pi} \right ) \W_{1}^{\star} \cdot \frac{\ru}{r^2}
\X \W_{1}\nonumber\\
&&- \frac{1}{2} \int d^{3} x (\W_{0}^{\star}  \X \W _{1} + \W_{1}^{\star}  \X \W_{0}
+ \W_{1}^{\star}  \X \W_{1})^2\nonumber\\
\eea
Also
\bea \label{5.11}
\int d^{3} x ~ |\D \X \W|^2 &=& \int d^{3} x ~ \W^{\star} \cdot \D \X (\D \X \W)
\nonumber\\
&=&\int d^{3} x ~ \left (
\W_{0}^{\star} \cdot \D \X (\D \X \W_{0}) 
+ \W_{1}^{\star} \cdot \D \X (\D \X \W_{1}) \right )
\eea
This is because the $\D \X$ operation does not change the $J,M$ values when
acting on $\clqJM$ and therefore  there are no terms mixing $\W_{0}$ and $\W_{1}$.
We have,
\bea 
\int d^{3} x ~ \W_{0}^{\star} \cdot \D \X (\D \X \W_{0})
= - \int dr w^{*} (r) \frac{d^2}{dr^2} w(r)
\eea
We will include the anomalous magnetic moment term
$-i  \W_{1}^{\star}\cdot (\ru /r^2) \X \W _{1}$ from Eq. (\ref{5.10})
along with the  $\W_{1}$ terms in Eq.\ (\ref{5.11}). Using the eigenfunctions
of the Hamiltonian $H$ given in Eq.\ (\ref{h})
\bea 
\int d^{3} x ~ \W_{1}^{\star}\cdot  H \W _{1} =
 \aJM \int_{-\infty}^{\infty} dk k^2 | \caqJM (k)|^2
\eea
 Thus finally,
\bea 
\label{final}
S& =& \int dr  \left[ \frac{2 \pi}{r^2} \left( 1 - \frac{|w(r)|^2}{4 \pi}\right)^2
- w^{*} (r) \frac{d^2}{dr^2} w (r) \right] 
+\aJM \int_{-\infty}^{\infty} dk\, k^2 | \caqJM (k)|^2 \nonumber\\
&&+ \frac{1}{2} \int d^{3}x (\nabla \X \abf)^2
+\int dr\frac{|w(r)|^2}{4\pi r^2}\sum_{J,M}(|F{^+_{qJM}}(r)|^2-|F{^-_{qJM}}(r)|^2)
\nonumber\\
&& - \frac{1}{2} \int d^{3} x (\W_{0}^{\star}  \X \W_{1}
+ \W_{1}^{\star}  \X \W_{0} + \W_{1}^{\star}  \X \W_{1})^2  \nonumber \\
&&+ i\int d^3x (\nabla \X \abf )\cdot (\W_{0}^{\star}  \X \W _{1} + \W_{1}^{\star}  \X \W _{0}
+ \W_{1}^{\star}  \X \W _{1})  \nonumber \\
&& +i \int d^3x\,\abf \cdot (\W^{\star} \X  (\D \X \W )-\W\X(\D \X \W)^{\star})\nonumber\\
&&+\int d^3x\,| \abf \X \W |^2\,.
\eea
Here
$
F{^\pm_{qJM}}(r)=\sum_\alpha\int dk\,\phi{^\alpha_{qJM}}(k)f{^{\pm\alpha}_{qJM}}(kr)
$,
with $F{^-_{qJM}}(r)=0$ for $J=1$. 

In Sec.\ \ref{ker}  we showed that the eigenfunctions with non-zero 
energy satisfy the constraint $\D \cdot \W=0$. Even though we are
not using the eigenfunctions of $H$ for the $J=0$ mode, this condition is valid
for this case too:
\bea 
\D \cdot\W_{0}
= \frac{dw(r)}{dr} \ru \cdot {\mathbf C^{+}_{100}} 
+  w(r) \D \cdot {\mathbf C^{+}_{100}}=0\,.
\eea
Therefore the ``natural gauge condition" \cite{p} for $\W$ is $\D \cdot \W =0$.
For the `photon' $\abf$, we may simply choose the gauge condition
$\nabla \cdot \abf=0$.

The functional measure $D\W^{*} D\W$is also simply transformed. Because of the 
orthonormality of the basis in Eqs.\ (\ref{wo}) and (\ref{w1}),
\bea
\int D\W^{\star} ~  D\W = \prod_{J,M,\alpha} \int d\caqJM(k) \int Dw(r)\,.
\eea
This completes the expression of the functional integral (for the case
of one monopole) using the new variables. It is in the form where
perturbative calculations including the topological degree can be
carried out. The free energy of one monopole in the leading order
will be presented elsewhere.

\section{Conclusion} \label{conc}

In this paper, we have initiated a method for systematically including topological
degrees of freedom within the ambit of renormalized perturbation theory. We
have argued that conventional semi-classical techniques are inapplicable for 
the case of pure gauge theories. Given the
advantages and successes of renormalized perturbation theory,
it is obviously better to adapt it to include the topological 
degrees systematically. We have addressed the
Yang-Mills theory in three Euclidean dimensions
in this paper. We have obtained a well-defined separation of the topological and
``spin wave'' degrees of freedom motivated by a singular gauge. 
The picture is simple enough: 
stray magnetic fields distorting the spherically symmetric magnetic fields of
Dirac monopoles, and massless charged vector bosons scattering off the latter.
We have explicitly shown that the Dirac string does not contribute.
We have considered the case of one monopole in detail. The $J=0$ mode of the
vector boson interactions gives precisely the core to make the energy of the 
monopole (ultraviolet) finite. 
The radial
equations are dramatically simplified
exactly with the anomalous magnetic moment
$g=2$ as required by the Yang-Mills symmetry. Also one polarization state of the
vector boson decouples.
We can evaluate the free energy of one monopole in the functional integral
by formally setting the gauge coupling constant $e$ to zero, and evaluate corrections to it 
in perturbation theory. It is to be noted that due to the Dirac quantization condition on the
monopole charge, the effects of the monopole does not disappear even when $e=0$.
This calculation and the effects of a gas of (anti-)monopoles  will be presented elsewhere.

\end{document}